\documentclass[prl,twocolumn]{revtex4-2}
\usepackage{dcolumn}
\usepackage{bm}
\usepackage{graphicx}
\usepackage{amsfonts}
\usepackage{bbm}
\usepackage{dsfont}
\usepackage{float}
\usepackage{graphicx}
\usepackage{amssymb}
\usepackage{multibib}
\usepackage{color}
\usepackage{enumerate}
\usepackage{amsmath}
\usepackage{amstext}
\usepackage{latexsym}
\usepackage{braket}
\usepackage{appendix}
\usepackage{textcomp}
\usepackage[usenames,dvipsnames]{xcolor}
\usepackage[colorlinks=true,citecolor=Blue,linkcolor=RubineRed,urlcolor=Blue]{hyperref}

\newcommand{\id}{\mathbf{1}_N}

\begin{document}
	
\title{Quantum-to-Classical Transition via Single-Shot Generalized Measurements}

\author{Zhenyu Xu\href{https://orcid.org/0000-0003-1049-6700}
{\includegraphics[scale=0.05]{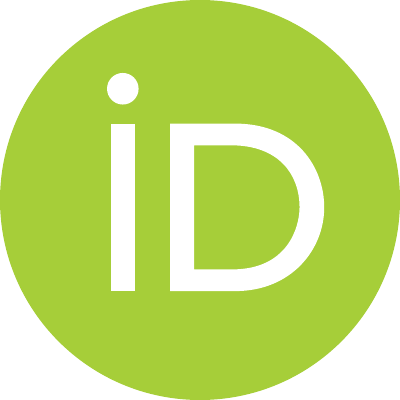}}}
\email{Contact author: zhenyuxu@suda.edu.cn}
\affiliation{School of Physical Science and Technology, Soochow University, Suzhou 215006, China}

\begin{abstract}
	We establish an operational connection between discrete rounds of generalized measurements and continuous-time decoherence, with an explicit correspondence between the number of measurement rounds and the evolution time. Operationally, we show that a single round of such a generalized measurement eliminates quasiprobability negativity in finite-dimensional systems. From the decoherence perspective, this loss of negativity occurs abruptly at a critical time. In particular, this critical time can be shorter than the conventional decoherence time, indicating that the latter does not always faithfully track the disappearance of nonclassicality. Our results provide new insight into the quantum-to-classical transition in finite-dimensional systems from the viewpoint of phase-space quasiprobability, and suggest feasible quantum-circuit tests as well as possible heralded resource extraction from noise.
\end{abstract}
	
\maketitle
		
\textit{Introduction}--Decoherence and measurement exhibit a subtle interplay in explaining the emergence of the classical world from the underlying quantum reality \cite{Zurek2003,Joos2003,Schlosshauer2005,Schlosshauer2019,Zurek_2025}. To explore the quantum-to-classical transition, in addition to descriptions based on the suppression of coherence in preferred bases \cite{Zurek2003,Joos2003,Schlosshauer2005,Schlosshauer2019,Zurek_2025} and the decay of quantum correlations \cite{SDE2009,Discord2012}, another natural approach is to work in quantum phase space \cite{bookPS}. In this framework, phase-space functions evolve according to a generalized Liouville equation, establishing direct analogies with classical dynamics \cite{bookPS}. Indeed, quantum phase-space formulations continue to find broad applications across quantum foundations \cite{Tilma2016PRL,ZhuHJ2016prl,Deffner2017NJP,Shanahan2018,Jeannic2018,Runeson2019,Oliva2019,Bohmann2020,Runeson2021PRL,Pedernales2023,Meng2023PRA,Xu2024PRL,Jorquera2025,Jarzynski2025,shrestha2026}, quantum measurement \cite{Descamps2023,Brody2025PRL,brody2025}, and quantum chaos \cite{Wang2023,Basu2024,Pizzi2025,Henning2025,Huh2025}, among others.

The emergence of negative values in quasiprobability phase-space functions is widely regarded as a witness of nonclassicality, since such features have no counterpart in classical physics \cite{Zurek2003,Joos2003,Schlosshauer2005,Schlosshauer2019,Zurek_2025,bookPS}. Recent advances in quantum technologies have renewed interest in phase-space representations of finite-dimensional systems, such as qu$d$its \cite{Rundle2021}. Although the quantum-to-classical transition in finite-dimensional systems has been extensively studied from the perspective of quantum correlations \cite{SDE2009,Discord2012}, the loss of quasiprobability negativity during this transition is still an open question, and the mechanism governing the intermediate regime remains unresolved.

In this Letter, we address this problem by establishing an explicit operational connection between discrete rounds of generalized $N$-level coherent-state positive-operator-valued measures (POVMs) and a continuous time open system dynamics. In particular, we derive a closed-form correspondence [Eq. \eqref{Dengjia}] that specifies when and how repeated generalized measurements reproduce the associated open-system dynamics. From the measurement viewpoint, we show that a single round of such a generalized measurement suffices to eliminate quasiprobability negativity in finite-dimensional systems. From the decoherence viewpoint, quasiprobability negativity disappears abruptly rather than decaying asymptotically. Moreover, the critical time for the loss of quasiprobability negativity can be shorter than the conventional decoherence time, showing that the latter can substantially overestimate how long phase-space nonclassicality persists in finite-dimensional quantum-to-classical transitions. Finally, we analyze the experimental feasibility of our protocol on current platforms, such as superconducting circuits, and discuss the possibility of extracting resources, including the preparation of $N$-level coherent states, from noisy channels.

\textit{$N$-level coherent-state POVM}--We start with the definitions. Without loss of generality, we consider a qu$d$it, that is, a quantum system with $N$ distinguishable levels ($d=N$). The generalized $N$-level coherent-state $|\Omega\rangle$ is constructed by applying a unitary operator $U_N(\Omega)=\prod_{k=1}^{N-1} R_{k, k+1}\left(\theta_{k}, \phi_{k}\right)$ to the ground state $|0\rangle_N$, namely $|\Omega\rangle = U_N(\Omega) |0\rangle_N$ (or constructed recursively; see, for example, Ref. \cite{Runeson2020}). Here, $\Omega$ is parameterized by $2N - 2$ angles, with $\theta_k \in[0, \pi]$ and $\phi_k \in[0,2 \pi)$ ($k=1,2,\cdots, N-1$), representing a generalization of the Bloch sphere. Each rotation gate $R_{k, k+1}(\theta_{k}, \phi_{k})=R_z(\phi_{k})R_x(\theta_k)R_z(-\phi_k)$ acts only on the subspace spanned by $\{|k\rangle,|k+1\rangle\}$, where $R_j(\alpha) = \exp(-i \alpha \sigma_j / 2)$ ($\sigma_{j\in \{x,y,z\}}$ denote the Pauli matrices).

In what follows, we consider the associated $N$-level coherent-state POVM \cite{POVM}, defined by
\begin{equation}
  \int_{\mathcal{M}} E(\Omega) \mathrm{d} \mu(\Omega)=\mathbf{1}_N, \label{povm}
\end{equation}
where $E(\Omega)=N|\Omega\rangle\langle\Omega|$ denotes a POVM element, and $\mathrm{d} \mu(\Omega)$ is the Haar measure on the manifold $\mathcal{M} = \mathrm{SU}(N)/\mathrm{U}(N-1) \cong \mathbb{C}\mathrm{P}^{N-1}$, satisfying $\int_{\mathcal{M}} \mathrm{d} \mu(\Omega) = 1$ \cite{Karol2001,Appleby2000,Kofler2008,Yang2014,Shojaee2018}. Since $\operatorname{Tr}[\rho E(\Omega)] = N \langle \Omega | \rho | \Omega \rangle \geq 0$ for any state $\rho$, the condition in Eq. (\ref{povm}) ensures that $E(\Omega)$ forms a valid POVM \cite{Davies1970,NielsenChuang2010,Uola2023RMP}, defined as a set of positive semidefinite Hermitian operators that sum to the identity ($\mathbf{1}_N$). 

The initial density operator $\rho_0$ evolves under one round of an $N$-level coherent-state POVM as $\mathcal{E}(\rho_0)=\int_{\mathcal M} \mathrm{d} \mu(\Omega) M(\Omega) \rho_0 M(\Omega)^{\dagger}$, where $M(\Omega)=\sqrt{E(\Omega)}=\sqrt{N}|\Omega\rangle\langle\Omega|$ denotes the corresponding measurement operator. This operation yields
\begin{equation}
  \mathcal{E}(\rho_0) = \frac{\mathbf{1}_N}{N} + \frac{\rho_0 - \mathbf{1}_N/N}{1+N}. \label{R1}
\end{equation}

\textit{Proof}.—Let $X$, $Y$, and $A$ act on the $N$-dimensional Hilbert space $\mathcal{H}$. 
Using the identity $\operatorname{Tr}_2[(\mathbf{1}_N \otimes A)(X \otimes Y)] = X\,\operatorname{Tr}(A Y)$ \cite{NoteI}, 
where $\operatorname{Tr}_2[\cdot]$ denotes the partial trace over the second subsystem of $\mathcal{H}^{\otimes 2}$, 
we set $A = \rho_0$ and $X = Y = |\Omega\rangle\langle\Omega|$ to obtain
\begin{equation}
\mathcal{E}(\rho_0)
  = N\, \operatorname{Tr}_2\!\left[
      (\mathbf{1}_N \otimes \rho_0)
      \int_{\mathcal{M}} \mathrm{d}\mu(\Omega)\,
      |\Omega\rangle\langle\Omega|^{\otimes 2}
    \right].
\end{equation}
This integral can be evaluated using the Haar second-moment identity $\int_{\mathcal{M}} \mathrm{d}\mu(\Omega)\,|\Omega\rangle\langle\Omega|^{\otimes 2} = (\mathbf{1}_N^{\otimes 2} + \mathsf{S})/[N(N+1)]$ \cite{collins2006integration}, 
where $\mathsf{S}$ swaps the two tensor factors, i.e., $\mathsf{S}(|\phi\rangle \otimes |\psi\rangle) = |\psi\rangle \otimes |\phi\rangle$. 
Substituting this relation gives 
$\mathcal{E}(\rho_0) = (1+N)^{-1} \operatorname{Tr}_2[(\mathbf{1}_N \otimes \rho_0)(\mathbf{1}_N^{\otimes 2} + \mathsf{S})] 
= (1+N)^{-1}(\mathbf{1}_N + \rho_0)$, 
which is equivalent to Eq. (\ref{R1}). (See Ref. \cite{SM} for an alternative derivation.)
\hfill $\blacksquare$

Thus, after $n$ rounds of POVMs, the initial state evolves into
\begin{equation}
  \rho_n=\underbrace{\mathcal{E} \circ \mathcal{E} \cdots \mathcal{E}}_n\left(\rho_0\right)=\frac{\mathbf{1}_N}{N}+\left(\frac{1}{1+N}\right)^n\left(\rho_0-\frac{\mathbf{1}_N}{N}\right). \label{rho_n}
\end{equation}
Clearly, as the number of measurement rounds increases, the state asymptotically approaches the maximally mixed state, independent of the initial condition. For instance, the Hilbert-Schmidt distance satisfies $\|\rho_n - \mathbf{1}_N / N\|_2 \propto (1 + N)^{-n}$, where $\|\cdot\|_2$ denotes the Hilbert-Schmidt norm. Thus every initial state converges toward the maximally mixed state as the number of measurement rounds increases. Such behavior is consistent with a depolarizing channel in the context of decoherence theory \cite{NielsenChuang2010}. In what follows, we justify this equivalence.

\textit{Generalized isotropic depolarizing channel}--The Gorini-Kossakowski-Sudarshan-Lindblad (GKSL) equation, also commonly referred to as the Lindblad master equation, can be written as ($\hbar=1$) \cite{GKS,Lindblad1976,BreuerBook,Rivas2012}
\begin{equation}
  \frac{\partial \rho_t}{\partial t}=-i\left[H, \rho_t\right]+\sum_l \gamma_l\left(L_l \rho_t L_l^{\dagger}-\frac{1}{2}\left\{L_l^{\dagger} L_l, \rho_t\right\}\right),
\end{equation}
where $\gamma_l$ are a set of non-negative coefficients representing decoherence rates, and $L_l$ are the Lindblad operators. Let $T_{\nu}$ $(\nu = 1, 2, 3, \ldots, N^2 - 1)$ be an orthonormal set of traceless Hermitian matrices that serve as the generators of the $\mathfrak{su}(N)$ Lie algebra \cite{Hall2015,Haber2021}, satisfying $\operatorname{Tr}\!\left(T_\mu T_\nu\right) = \delta_{\mu \nu}/2$. We now introduce a generalized isotropic depolarizing channel by setting $ L_\nu = T_\nu $ with a uniform decoherence rate $\gamma_\nu \equiv \gamma $, which leads to 
\begin{equation}
  \frac{\partial \rho_t}{\partial t}=\gamma \left(T_\nu \rho_t T_\nu-\frac{N^2-1}{2 N}\rho_t\right)=\frac{N \gamma}{2} \left(\frac{\mathbf{1}_N}{N}-\rho_t \right), \label{GIDC}
\end{equation}
where we have used $T_\nu T_\nu=\frac{N^2-1}{2N} \mathbf{1}_N$ \cite{Haber2021} and the summation over repeated Greek indices is implicit. For the second equality in Eq. (\ref{GIDC}), we have employed $T_\nu X T_\nu=\operatorname{Tr}(X) \mathbf{1}_N/2-X/(2N)$ \cite{Haber2021}. The solution for Eq. (\ref{GIDC}) is straightforward 
\begin{equation}
  \rho_t=\frac{\mathbf{1}_N}{N}+e^{-\frac{N}{2} \gamma t}\left(\rho_0-\frac{\mathbf{1}_N}{N}\right). \label{rhott}
\end{equation}
By comparing Eq. (\ref{rhott}) with Eq. (\ref{rho_n}), we find that setting
\begin{equation}
    t=\frac{2n}{\gamma N}\ln(1+N),\label{Dengjia}
\end{equation}
yields an evolved state $\rho_t$ that exactly matches $\rho_n$ \cite{notet}.

\textit{Quantum-to-classical transition in phase space}--For finite-dimensional quantum systems, the phase-space formalism was originally introduced by Stratonovich in a continuous formulation \cite{Stratonovich1957}, and later developed in alternative formulations, including discrete phase-space approaches \cite{WOOTTERS19871,GALETTI1999473}. Both the continuous and discrete formalisms have since undergone substantial development, particularly with the rise of quantum information and quantum technologies \cite{Rundle2021}. In this Letter, we focus on the continuous formulation (the discrete phase-space case is briefly discussed in Ref. \cite{SM}). The key idea is to map an operator $A$ in Hilbert space to an $s$-parametrized quasiprobability distribution function $W_A^{(s)}(\Omega)$ in the $\Omega$-spanned phase space via the Stratonovich-Weyl (SW) kernel $\Delta^{(s)}(\Omega)$, namely, 
\begin{equation}
  W_A^{(s)}(\Omega)=\operatorname{Tr}\left[A \Delta^{(s)}(\Omega)\right]. \label{PPF}
\end{equation} 
For $N$-level quantum systems, such SW kernel can be constructed as $\Delta^{(s)}(\Omega)=\mathbf{1}_N/N+2 r_s R_\nu T_\nu$, where $R_\nu=\langle\Omega| T_\nu|\Omega\rangle$, and $r_s=(1+N)^{\frac{1+s}{2}}$ is the $s$-parametrized radius \cite{Runeson2020}. This SW kernel satisfies the five fundamental criteria: linearity, reality, standardization, covariance, and trace preservation \cite{Meng2023PRA}. These properties ensure a valid and self-consistent SW correspondence \cite{Rundle2021}. Moreover, this kernel naturally parallels the continuous-variable Cahill-Glauber $s$-ordering formalism \cite{Glauber1969,Cahill1969}, yielding the Wigner, Glauber-Sudarshan $P$, and Husimi $Q$ quasiprobability distributions for $s=0$, $1$, and $-1$, respectively. 

\begin{figure}[t]
	\centering
	\includegraphics[width=8.5cm]{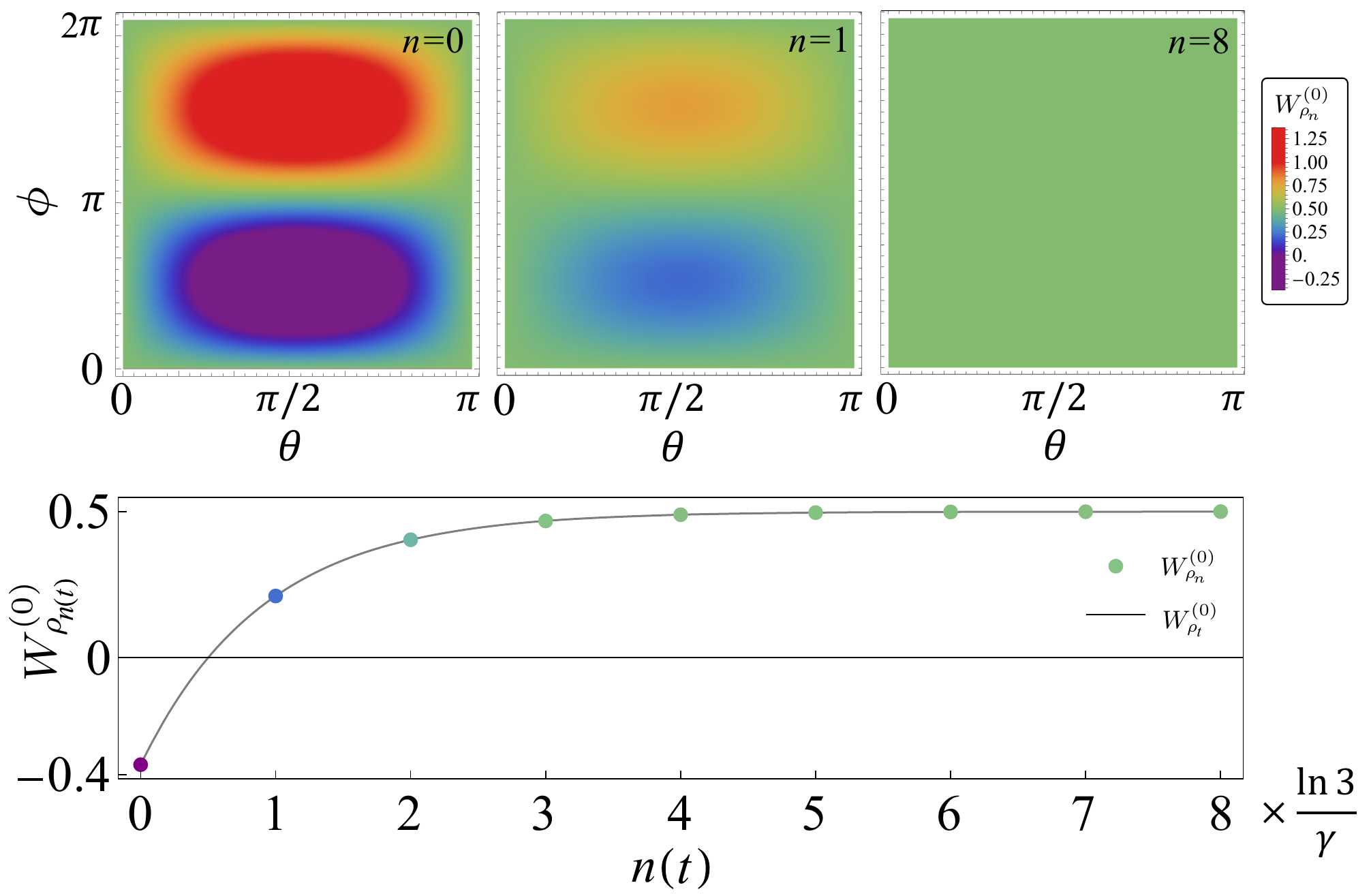}
	\caption{\textbf{ Loss of Wigner negativity via a single round of POVM.} Top: The Wigner quasiprobability distribution $W_{\rho_n}^{(0)}$ for a qubit after $n$ rounds of two-level coherent-state POVMs. The coherent state is given by $|\Omega\rangle = \cos (\theta/2)|0\rangle + \sin (\theta/2)\,\mathrm{e}^{\mathrm{i}\phi}|1\rangle$. Bottom: Equivalence between the coherent-state POVM and the isotropic depolarizing channel. The curve and markers (for $\theta=\phi=\pi/2$) illustrate the process of the disappearance of Wigner negativity from both viewpoints.}
	\label{fig_qubit}
\end{figure}

Equation (\ref{rho_n}) or (\ref{rhott}) now in the phase space reads 
\begin{equation}
  W^{(s)}_{\rho_{n(t)}}(\Omega)=\frac{1}{N}+\Gamma_{n(t)}\left(W^{(s)}_{\rho_0}(\Omega)-\frac{1}{N}\right), \label{Wt}
\end{equation}
with $\Gamma_n=1/(1+N)^n $ and $\Gamma_t=e^{-\frac{N}{2} \gamma t}$, respectively. To this end, we first examine the phase-space function corresponding to an arbitrary initial state $\rho_0$, yielding
\begin{equation}
  W^{(s)}_{\rho_0}(\Omega)=\frac{1-r_s}{N}+r_s\langle\Omega| \rho_0|\Omega\rangle,\label{W0}
\end{equation}
where we have employed $|\Omega\rangle\langle\Omega|=\mathbf{1}_N / N+2 R_\nu T_\nu$. Since $\langle\Omega|\rho_0|\Omega\rangle \ge 0$, the lower bound is saturated when the initial state is pure and orthogonal to $|\Omega\rangle$. 
In this case, $W^{(s)}_{\rho_0}(\Omega) \ge (1 - r_s)/N \ge -1$, where the second inequality follows by restricting the Cahill-Glauber ordering parameter to $s \in [-1, 1]$ \cite{Runeson2020}. Therefore, according to Eq. (\ref{Wt}), $W^{(s)}_{\rho_1}(\Omega) = \frac{1 + W^{(s)}_{\rho_0}(\Omega)}{N + 1} \ge 0$. This result shows that a single round of coherent-state POVM removes all negativity in the phase-space quasiprobability, regardless of the initial state or the system dimension $N$.

As an example, consider the widely used Wigner phase space ($s = 0$) and the Glauber–Sudarshan $P$ function ($s = 1$). In the extreme case where the initial state is pure and orthogonal to $|\Omega\rangle$, one round of POVM transforms the initially minimal negative value $W^{(0)}_{\rho_{0}}(\Omega) = (1 - \sqrt{N + 1}) / N < 0$ into a positive one $W^{(0)}_{\rho_{1}}(\Omega) = (1 - 1 / \sqrt{N + 1}) / N > 0$, and likewise $W^{(1)}_{\rho_{0}}(\Omega) = -1$ into $W^{(1)}_{\rho_{1}}(\Omega) = 0$ (non-negative) \cite{NoteWigner}.

For illustration, we consider the Wigner phase space and $N = 2$, i.e., a qubit, in Fig. \ref{fig_qubit}. The initial state is chosen to display minimal negativity. As seen in the top panels of Fig. \ref{fig_qubit}, the purple region, which represents the negative values of the phase-space function, vanishes after a single round of the generalized measurement. From the viewpoint of decoherence, this behaviour is equivalent to the action of an isotropic depolarizing channel. The bottom panel of Fig. \ref{fig_qubit} shows the intermediate stages of the disappearance of the Wigner negativity from both perspectives.

The loss of Wigner negativity induced by the above single-shot generalized measurement in finite-dimensional systems has a natural analogue in continuous-variable systems. By Hudson’s theorem, the Wigner function of a continuous-variable system is non-negative if and only if it is Gaussian in phase space \cite{HUDSON1974249}. A phase-space measurement onto coherent states yields conditioned post-measurement states with Gaussian Wigner functions, thereby removing Wigner negativity, see, e.g., Ref. \cite{Brody2025PRL}.

\begin{figure}[b]
	\centering
	\includegraphics[width=7.5cm]{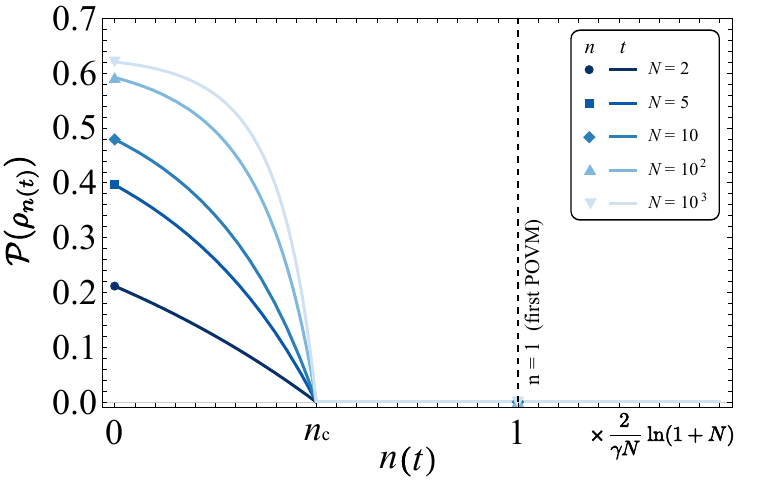}
	\caption{\textbf{Sudden vanishing of the negative Wigner quasiprobability volume.} The volume of the negative Wigner quasiprobability function, $\mathcal{P}(\rho_{n(t)})$, is plotted against time $t$ or the number of POVM rounds $n$, related by $t = \tfrac{2n}{\gamma N}\ln(1 + N)$ [Eq. (\ref{Dengjia})]. Solid curves $\mathcal{P}(\rho_t)$ correspond to $N = 2, 5, 10, 10^{2}, 10^{3}$ (dark to light blue) and depict the evolution of the negative-quasiprobability volume over time, showing a sudden disappearance at $t_c (n_c = 1/2)$. Markers at $n = 0$ denote the initial negative-quasiprobability volume, while those at $n = 1$ (dashed vertical line) indicate that a single coherent-state POVM completely eliminates the negativity.
}
	\label{fig_negativeP}
\end{figure}

\textit{Sudden vanishing of negative quasiprobability volume.}--The above analysis can also be formulated in terms of the phase-space volume of the negative quasiprobability region:
\begin{equation}
 \mathcal{P}(\rho_{n(t)})=\int_{\mathcal{M}^\star} \mathrm{d}\mu (\Omega) \quad \text{with} \quad \mathcal{M}^\star = \{\Omega | W^{(s)}_{\rho_{n(t)}}<0\}. \label{NPro}
\end{equation}
We first consider Eq. (\ref{NPro}) for the initial state $\rho_0$. According to Eq. (\ref{W0}), we define the negativity threshold as 
$p_c = (r_s - 1) / (N r_s)$, 
so that $\mathcal{M}^\star = \{\Omega \,|\, \langle\Omega|\rho_0|\Omega\rangle \le p_c\}$. 
Let $\rho_0 = \sum_{j=1}^N \lambda_j |e_j\rangle\!\langle e_j|$ with eigenvalues 
$\lambda_1 \le \cdots \le \lambda_N$. 
After some algebra \cite{SM,deBoor2001}, we obtain  
\begin{equation}
  \mathcal{P}(\rho_0) = 
  \frac{1}{(N - 1)!} 
  \sum_{\lambda_j < p_c} 
  \frac{(p_c - \lambda_j)^{N - 1}}{\prod_{k \neq j} (\lambda_j - \lambda_k)}.
  \label{Pro}
\end{equation}
The summation runs only over eigenvalues strictly below $p_c$. 
For the special case of an initial pure state, Eq. (\ref{Pro}) reduces to 
$\mathcal{P}(\rho_0) = 1 - (1 - p_c)^{N - 1}$. In contrast, $\mathcal{P}(\rho_1) \equiv 0$. This result is consistent with the previous analysis, confirming that a single round of coherent-state POVM completely eliminates the negativity of the phase-space function.

From the perspective of decoherence, similar calculations show that for $0 \le t \le t_c$, $\mathcal{P}(\rho_t)$ takes the same form as Eq. (\ref{Pro}), with $p_c$ replaced by $p_c(t) = [1 - 1/(\Gamma_t r_s)] / N$. 
Here, $t_c$ denotes the critical time, beyond which the negative quasiprobability volume vanishes \cite{SD} (see Fig. \ref{fig_negativeP}). For the Wigner function, the critical time for the abrupt loss of Wigner negativity is given by \cite{SM}
\begin{equation}
    t_c=\frac{2}{\gamma N} \ln \left[\sqrt{N+1}\left(1-N \lambda_{\min }\right)\right],
    \label{tc}
\end{equation}
where $\lambda_{\min}$ denotes the smallest eigenvalue of $\rho_0$, equivalently $\lambda_{\min}=\min_{\Omega}\langle\Omega|\rho_0|\Omega\rangle$ (for a pure initial state, $\lambda_{\min}=0$).

\begin{figure}[t]
	\centering
	\includegraphics[width=7.5cm]{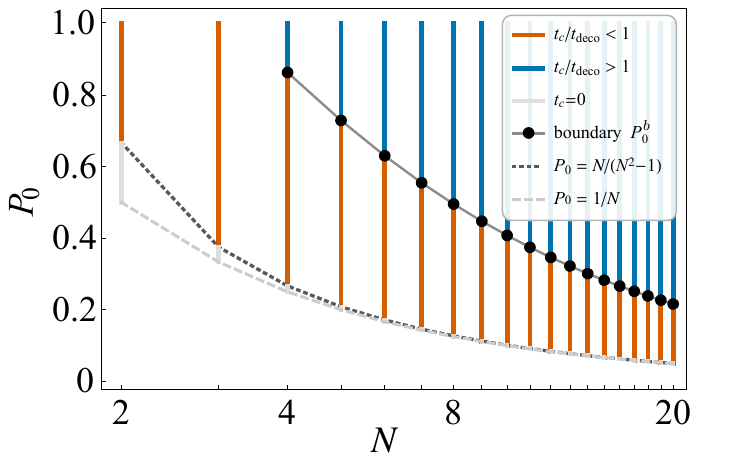}
	\caption{\textbf{Discrete phase diagram in the $(N,P_0)$ plane for the ratio between the critical time $t_c$ and the decoherence time $t_{\mathrm{deco}}$.} 
For each Hilbert-space dimension $N \in [2,20]$, colored vertical segments indicate the range of initial purities $P_0$: red denotes purities for which there exists an initial state $\rho_0$ exhibiting Wigner negativity and satisfying $t_c/t_{\mathrm{deco}}<1$, while blue denotes the complementary regime $t_c/t_{\mathrm{deco}}>1$. Light gray segments correspond to states without initial Wigner negativity ($t_c=0$). Black circles mark the boundary purity $P_0^b$ defined by $\big(t_c/t_{\mathrm{deco}}\big)_{\min}=1$. The dotted curve shows the negativity existence threshold $P_0=N/(N^2-1)$, and the dashed curve is the purity lower bound $P_0=1/N$.
}
	\label{fig_tctd}
\end{figure}

\textit{Critical time versus decoherence time}--Since the critical time $t_c$ can differ substantially from the conventional decoherence time, it is natural to compare the two. Following the work of Zurek and others, we employ the short-time asymptotic expansion of the purity, defined as $P_t=\mathrm{Tr}(\rho_t^2)$, namely, $P_t/P_0\simeq 1-t/t_{\mathrm{deco}}+\mathcal{O}(t^2)$, with $t_{\mathrm{deco}}=-P_0/[2\,\mathrm{Tr}\!\left(\rho_0\dot{\rho}_0\right)]$ \cite{Zurek2003,Lidar98,LIDAR200682,Bedingham2014,Chenu2017,Beau2017,Xu19}. Using Eq. \eqref{rhott}, this gives $t_{\mathrm{deco}}=P_0/[\gamma\left(N P_0-1\right)]$. Then, by means of Eq. \eqref{tc}, we obtain the ratio
\begin{equation}
   \frac{t_c}{t_\mathrm{deco}}= \frac{2(NP_0-1)}{NP_0}\ln\!\Big[\sqrt{N+1}\,(1-N\lambda_{\min})\Big],
   \label{ratio}
\end{equation}
which depends on the initial state, through the purity $P_0$ and $\lambda_{\min}$, as well as on the system dimension. In the large dimension limit ($N\to\infty$) and for a pure initial state, one finds $t_c/t_{\mathrm{deco}}\propto \ln N$. This implies that, in this scenario, the conventional definition of the decoherence time remains a reasonable indicator, as the decoherence time is shorter than the critical time. However, for finite dimension and an arbitrary initial (mixed) state, the situation becomes more subtle. For $N=2$ and $3$, the ratio is always less than unity (see Fig. \ref{fig_tctd}). For $N\geq 4$, one may define a boundary purity $P_0^b$, namely,
\begin{equation}
    P_0^b=\frac{1}{N}\left(1+\frac{1}{W\left(\frac{N+1}{e(N-1)}\right)}\right), \quad N \geq 4,
\label{Pstar}
\end{equation}
where $W(\cdot)$ denotes the Lambert $W$ function (principal branch). Then, for $N/(N^2-1)<P_0<P_0^b$, there exist initial states for which $t_c/t_{\mathrm{deco}}<1$ (as illustrated in Fig. \ref{fig_tctd}). This shows that the conventional decoherence time need not faithfully track the disappearance of Wigner negativity in this regime.

\textit{Experimental proposal and feasibility}--The above theory can be implemented on state-of-the-art experimental platforms, such as superconducting circuits \cite{Clarke2008,Huang2020,Kjaergaard2020,Jiang2025}. The central task is to evaluate the POVM statistics, i.e., $\operatorname{Tr}[\rho E(\Omega)]$, which can be implemented either probabilistically using an ancilla-free strategy or deterministically with an ancilla-assisted strategy \cite{DQC1,Poulin2003PRA,Swingle2016,Zoller2020PRXQ,Quan2006PRL,Zhang2008PRL,Dorner2013PRL,Mazzola2013PRL,Paternostro2014PRL,Wei12,Peng15,RN1221,Zhenyu2019PRLmanybody,Cai2019,Lu2024PRL,Lu2025PRL} (see details in Ref. \cite{SM}). For both implementations, the circuit depth scales linearly with the Hilbert-space dimension, $\Theta(N)$ \cite{LCU2012}. Let the fidelity of a single two-qubit gate be $f$ (with infidelity $r = 1 - f$) and the gate time be $\tau$. 
The circuit consists of $L$ two-qubit layers with $L \simeq \kappa N$, where $\kappa$ is a circuit-dependent constant. 
The total gate time of the circuit is therefore $T_{\mathrm{tot}} \simeq L \tau$, and the total fidelity is $F_{\mathrm{tot}} \simeq (1 - r)^L \simeq e^{-rL}$. 
If the minimum acceptable total fidelity is $F_{\min}$, the maximum implementable system size satisfies $ N \lesssim \min \left\{\frac{-\ln F_{\min }}{\kappa r}, \frac{T_2}{\kappa \tau}\right\}$. Current superconducting platforms report two-qubit gate fidelities of $99.7$–$99.9\%$, gate times of $20$–$160$ ns, and coherence times in the range $T_{2} \sim 0.3$–$0.5$ ms for transmons and $\gtrsim 1$ ms for fluxonium \cite{Clarke2008,Huang2020,Kjaergaard2020,Jiang2025}. 
For a conservative analysis, assuming $F_{\min}=95\%$, $T_2=0.3~\mathrm{ms}$, $\kappa=1$, $r=0.1\%$, and $\tau=100~\mathrm{ns}$, our estimates suggest that the protocol remains experimentally feasible up to approximately $N\lesssim51$, which is sufficient to observe the single-shot negativity-removal effect predicted in this Letter.

\textit{Conclusions and outlook}--We have connected discrete generalized measurements to continuous decoherence in an operational manner, providing new insight into the intermediate regime of the quantum-to-classical transition in finite-dimensional systems from the perspective of quasiprobability negativity in phase space. Within this framework, we show that a single round of such a generalized measurement eliminates quasiprobability negativity in finite-dimensional systems. From a decoherence viewpoint, this loss of quasiprobability negativity occurs abruptly at a critical time, which can be shorter than the conventional decoherence time.

The single-shot loss of Wigner negativity identified in this Letter suggests that analogous threshold behavior may also arise for other witnesses of quantumness, such as entanglement, and warrants further study. It is also worth noting that Wigner negativity is not only a phase-space signature of nonclassicality, but is also closely connected to the resource of ``magic'' underlying quantum computational advantage beyond stabilizer operations \cite{Kitaev2005,Veitch2012,Howard2014,Nicolas2015,qmw5-2wbf,5c15-4g5n,1tyr-rlbb}. From this resource-oriented perspective, our results connect the removal of phase-space negativity by single-shot measurements to the broader question of how efficiently magic monotones can be suppressed.

At the same time, although the induced decoherence map is detrimental at the ensemble level, monitoring the underlying measurement record may yield conditional, trajectory-level pure outputs, thereby opening the possibility of heralded preparation of useful states from noise; see Ref. \cite{SM} for a possible proposal and discussion. Developing such heralded resource-extraction protocols for realistic noise models and finite-outcome discretizations, as well as clarifying which resources can be generated, remains an open direction in the NISQ era \cite{Preskill2018}.

\textit{Acknowledgements}--This work was supported by the National Natural Science Foundation of China under Grants No. 12074280 and No. 12574398.


\bibliography{references,references_PhaseSpace,references_Chaos,refsRM}

\begin{thebibliography}{96}
\bibitem [{Not()}]{NoteI}%
  \BibitemOpen
  \href@noop {} {}\bibinfo {note} {Let $X,Y,A$ act on $\mathcal{H}$ and fix an orthonormal basis $\{\lvert r\rangle\}$ for the second tensor factor. Define $ F \;=\; \operatorname{Tr}_2\!\big[(\mathbbm{1}_N \otimes A)(X\otimes Y)\big]$. Then, for any $\lvert i\rangle$ and $\lvert j\rangle$, $\langle i \lvert F \rvert j \rangle=\sum_{r}\,\langle i,r \lvert (\mathbbm{1}_N\otimes A)(X\otimes Y) \rvert j,r \rangle=\sum_{r}\,\langle i \lvert X \rvert j \rangle \;\langle r \lvert A Y \rvert r \rangle=\langle i \lvert X \rvert j \rangle \;\operatorname{Tr}(AY).$ Since the equality holds for all matrix elements, which ends the proof of the identity used in the main text}\BibitemShut {NoStop}%
\bibitem [{Not()}]{NoteWigner}%
  \BibitemOpen
  \href@noop {} {}\bibinfo {note} {In the remainder of this Letter, we focus on the widely used Wigner function with $s=0$, unless explicitly stated otherwise}\BibitemShut {NoStop}%
\end{thebibliography}
	
\appendix\label{app}
\widetext
\begin{center}
\textbf{Supplemental Material}
\end{center}

\setcounter{equation}{0} \setcounter{figure}{0} \setcounter{table}{0}
\makeatletter
\renewcommand{\theequation}{S\arabic{equation}} \renewcommand{\thefigure}{S%
\arabic{figure}} \renewcommand{\bibnumfmt}[1]{[#1]} \renewcommand{%
\citenumfont}[1]{#1}

\tableofcontents


\section{I. An alternative derivation of Eq. (2) in the main text}\label{appendixA}
In the Letter, we use the Haar second-moment identity for the derivation of Eq. (2), which is concise but less accessible. In this section, we provide an alternative approach that is more pedagogical and easier to follow.

The measurement operator is defined by $M(\Omega)=\sqrt{E(\Omega)}=\sqrt{N}|\Omega\rangle\langle\Omega|$. Then, the density operator $\rho$ after applying one POVM is given by
\begin{equation}
  \mathcal{E}(\rho)=\int_{\mathcal M} \mathrm{d} \mu(\Omega) M(\Omega) \rho M(\Omega)^{\dagger}=\int_{\mathcal M} N|\Omega\rangle\langle\Omega| \rho|\Omega\rangle\langle\Omega| \mathrm{d} \mu(\Omega). \label{GPOVM}
\end{equation}
Assume $\rho=\frac{1}{N}\id+b_\nu T_\nu$, where repeated Greek indices are implicitly summed, and $b_\nu=2\,\mathrm{Tr}(\rho T_\nu)$. Here $T_\nu$ ($\nu=1,2,3,\dots,N^2-1$) is an orthonormal set of traceless Hermitian matrices
that form a basis of $\mathfrak{su}(N)$ (up to the conventional factor of $i$) Lie algebra, satisfying $\mathrm{Tr}(T_\mu T_\nu)=\delta_{\mu\nu}/2$ \cite{Haber2021}.
Since the POVM is covariant under $\text{SU}(N)$, the induced channel is unital and $\text{SU}(N)$-covariant, i.e.,
\begin{equation}
    \mathcal{E}(\id)=\id,\qquad
    \mathcal{E}(U\rho U^\dagger)=U\,\mathcal{E}(\rho)\,U^\dagger,\ \forall\,U\in \text{SU}(N).
    \label{S2}
\end{equation}
Therefore, $\mathcal{E}$ leaves the identity component invariant and maps the traceless subspace
$\mathfrak{su}(N)$ into itself. Since the adjoint representation of $\mathrm{SU}(N)$ on $\mathfrak{su}(N)$ is irreducible, Schur's lemma implies that $\mathcal{E}$ acts as a scalar on this subspace \cite{Hall2015}, namely
\begin{equation}
    \mathcal{E}(T_\nu)=\lambda\,T_\nu,
    \label{ETnu}
\end{equation}
with a constant $\lambda$ independent of $\nu$.
Using Eqs. \eqref{S2} and \eqref{ETnu}, we obtain
\begin{equation}
  \mathcal{E}(\rho)=\frac{1}{N}\id+\lambda\,b_\nu T_\nu.
\label{S4}
\end{equation}
To extract $\lambda$, we take the Hilbert-Schmidt inner product with $T_\nu$, i.e.,
\begin{equation}
    \mathrm{Tr}\!\left(T_\nu\mathcal{E}(T_\nu)\right)=\lambda\,\mathrm{Tr}(T_\nu T_\nu)=\frac{\lambda}{2}(N^2-1),
\end{equation}
where summation over $\nu$ (repeated Greek indices) is implicit hereafter. Equivalently, 
\begin{equation}
    \lambda=\frac{2}{N^2-1}\mathrm{Tr}\!\left(T_\nu\mathcal{E}(T_\nu)\right).
\label{Lambda0}
\end{equation}
Substituting Eq. (\ref{GPOVM}) into Eq. (\ref{Lambda0}), we have
\begin{equation}
  \lambda=\frac{2}{N^2-1} \int_{\mathcal{M}} N \operatorname{Tr}\left[T_\nu|\Omega\rangle\langle\Omega| T_\nu|\Omega\rangle\langle\Omega|\right] \mathrm{d} \mu(\Omega)=\frac{2 N}{N^2-1} \int_{\mathcal{M}} \langle\Omega| T_\nu|\Omega\rangle ^2 \mathrm{d} \mu(\Omega).\label{Lambda}
\end{equation}
Taking into account
\begin{equation}
    \langle\Omega|T_\nu|\Omega\rangle^2
    =\left[\mathrm{Tr}\!\left(|\Omega\rangle\langle\Omega|\,T_\nu\right)\right]^2
    =\frac{1}{4} b_\nu(\Omega) b_\nu(\Omega),
\end{equation}
where $b_\nu(\Omega)=2 \operatorname{Tr}\left(|\Omega\rangle\langle\Omega| T_\nu\right)=2\langle\Omega| T_\nu|\Omega\rangle,$ and using
\begin{equation}
    \mathrm{Tr}(|\Omega\rangle\langle\Omega|^2)
    =\frac{1}{N}+\frac{1}{2}b_\nu(\Omega)b_\nu(\Omega)=1, \qquad\text{i.e.,}\qquad
    b_\nu(\Omega)b_\nu(\Omega)=\frac{2(N-1)}{N},
\end{equation}
we obtain
\begin{equation}
    \langle\Omega|T_\nu|\Omega\rangle^2=\frac{1}{2}\!\left(1-\frac{1}{N}\right).
    \label{lambdaF}
\end{equation}
Equation \eqref{lambdaF} can also be derived alternatively; see, for example, Eq. (E3) of Ref. \cite{Runeson2020}. With Eqs. \eqref{Lambda} and \eqref{lambdaF}, we obtain
\begin{equation}
  \lambda=\frac{1}{N+1}.
\end{equation}
Thus, the initial state $\rho_0=\frac{1}{N} \id+b_{\nu}(0) T_{\nu}$ after one $N$-level coherent state POVM reads
\begin{equation}
  \mathcal{E}\left(\rho_0\right)=\frac{\id}{N}+\lambda b_\nu(0) T_\nu=\frac{\id}{N}+\frac{1}{N+1}\left(\rho_0-\frac{\id}{N}\right),
  \label{unital}
\end{equation}
which ends the alternative proof of Eq. (2) in the main text.

\section{II. Discrete phase space and single-shot positivity}
For finite-dimensional systems, phase-space descriptions may alternatively be formulated on a finite lattice, known as a discrete phase space \cite{WOOTTERS19871,GALETTI1999473}. For illustration, we consider a qubit and briefly show that a central conclusion of this work, namely that a single round of the coherent-state POVM removes Wigner negativity, remains valid in the discrete phase space.

For a qubit, we can label phase-space points by $\eta\equiv(a_1,a_2)$ with $a_{1,2}\in\{0,1\}$. A kernel for the discrete phase space can be written as
\begin{equation}
    \Delta(\eta)=\frac{1}{2} \mathbf{1}_2+\frac{1}{2}\left[(-1)^{a_2} \sigma_x+(-1)^{a_1+a_2} \sigma_y+(-1)^{a_1} \sigma_z\right],
\label{discretekernel}
\end{equation}
where $\mathbf{1}_2$ is the identity operator for a qubit, and $\sigma_j$ ($j=x,y,z$) are the usual Pauli operators. 

Write the initial qubit state as $\rho_{0}=\frac12(\mathbf{1}_2+\bm r\cdot\bm\sigma)$ with $|\bm r|\le 1$, and define the (fixed) lattice vectors
\begin{equation}
    \boldsymbol{n}(\eta)=\left((-1)^{a_2},(-1)^{a_1+a_2},(-1)^{a_1}\right), \quad|\boldsymbol{n}(\eta)|=\sqrt{3},
    \label{Lattice}
\end{equation}
one obtains the discrete phase space function for the initial state
\begin{equation}
    W_{\rho_0}(\eta)=\text{Tr}[\rho_0 \Delta(\eta)]=\frac12\bigl[1+\bm r\cdot\bm n(\eta)\bigr].
    \label{stateOW}
\end{equation}
Hence the most negative value over all qubit states is achieved by a pure state antiparallel to $\bm n(\eta)$, giving
$W_{\min}=\frac12(1-\sqrt3)<0$.

Now apply one round of our coherent-state POVM. In the main text, the induced channel is given by $\rho_1=\mathcal{E}\left(\rho_0\right)=\mathbf{1}_2/2+(\rho_0-\mathbf{1}_2 / 2)/3$, which gives 
\begin{equation}
    W_{\rho_1}(\eta)=\frac{W_{\rho_0}(\eta)+1}{3}.
    \label{rho1}
\end{equation}
Using $|\bm r\cdot\bm n(\eta)|\le |\bm r|\,|\bm n(\eta)|\le \sqrt3$, we obtain the uniform lower bound
\begin{equation}
W_{\rho_1}(\eta)\ge \frac12\!\left(1-\frac{\sqrt3}{3}\right)>0\qquad\forall\,\eta,
\label{eq:discrete_W_positive_after_one}
\end{equation}
for any initial qubit state $\rho_0$. Thus, even in the discrete qubit phase space, a single-shot coherent-state POVM suffices to remove all Wigner negativity, consistent with the continuous phase space analyzed in the main text.

\section{III. Decoherence time versus the finite vanishing time of Wigner negativity}
\subsection{a. Decoherence time $t_\mathrm{deco}$ of general Markovian dynamics}
We consider the evolved quantum state $\rho_t$ governed by the Markovian master equation in the Lindblad form ($\hbar=1$)
\begin{equation}
\dot{\rho}_{t}=-i\left[ H,\rho _{t}\right] +\sum_{l}\gamma
_{l}\left( L_{l}\rho _{t}L_{l}^{\dag }-\frac{1}{2}L_{l}^{\dag }L_{l}\rho
_{t}-\frac{1}{2}\rho _{t}L_{l}^{\dag }L_{l}\right) ,  \label{Master Eq}
\end{equation}%
where the Hamiltonian $H$ is related to the open system, the non-negative
quantities $\gamma _{l}$ play the role of relaxation rates and $L_{l}$ are
the Lindblad operators.

To quantify how rapidly an open system decoheres, we first define a measure of the mixedness of a quantum state induced by its interaction with the environment. Here we use the purity, defined as
\begin{equation}
    P_{t}=\text{Tr}\left( \rho _{t}^{2}\right) .  \label{purity}
\end{equation}
To identify a universal expression of the decoherence time under Markovian
evolution, we could explore the short-time asymptotic behavior of quantum
dynamics by expanding $\rho _{t}$ up to the order of $t$, i.e., $\rho
_{t}\simeq \rho _{0}+\dot{\rho}_{0}t+\mathcal{O}(t^{2})$. Then, the
short-time asymptotic expansion of Eq. (\ref{purity}) reads \cite{Zurek2003,Lidar98}
\begin{equation}
    P_{t}/P_0\simeq 1-t/t_\mathrm{deco}+\mathcal{O}(t^{2}),  \label{pt}
\end{equation}
where the decoherence time $t_\mathrm{deco}$ is defined as 
\begin{equation}
    t_\mathrm{deco}=-\frac{P_0 }{2\text{Tr}\left( \rho _{0}\dot{\rho}_{0}\right) }=\frac{P_0 }{2\sum_{l}\gamma _{l}\widetilde{\mathrm{cov}}_{\rho _{0}}(L_{l}^{\dag },L_{l})}.
\label{dec-time}
\end{equation}
Here $\widetilde{\mathrm{cov}}_{\rho
_{0}}(X,Y)=\left\langle \rho _{0}XY\right\rangle _{\rho _{0}}-\left\langle
X\rho _{0}Y\right\rangle _{\rho _{0}}$ is the modified covariance, with $\left\langle X\right\rangle_{\rho _{0}}=$Tr$\left( \rho _{0}X\right) $ (similar expansions based on the fidelity or survival amplitude can be found in Refs. \cite{Chenu2017,Beau2017}). Obviously, the decoherence time has a clear geometric interpretation, since it is directly related to the slope of purity in the short-time scale. 

We further note that higher-order expansions in time $t$ may also be considered. In this Letter, however, we restrict attention to unital dynamics, for which all Taylor coefficients are governed by the same decoherence rate. Consequently, higher-order coefficients are not independent. This point will be explained in more detail in the following subsection.

\subsection{b. Decoherence time $t_\mathrm{deco}$ of unital dynamics in the main text}
In the main text, we consider the Lindblad operators sampled from the generators of the $\mathfrak{su}(N)$ Lie algebra \cite{Haber2021}, i.e., $ L_\nu = T_\nu $ $(\nu = 1, 2, 3, \ldots, N^2 - 1)$. In addition, we assume the decoherence rate to be uniform, $\gamma_\nu\equiv\gamma$. Equation (\ref{dec-time}) now reads
\begin{equation}
    t_\mathrm{deco}=\frac{P_0}{2 \gamma\left[\operatorname{Tr}\left(\rho_0^2 T_\nu T_\nu\right)-\operatorname{Tr}\left(\rho_0 T_\nu \rho_0 T_\nu\right)\right]}, 
    \label{dec-time1}
\end{equation}
where the summation over repeated Greek indices is again implicit. By employing the identity $T_\nu T_\nu=\frac{N^2-1}{2N} \id$ \cite{Haber2021}, the first summation in the denominator of Eq. (\ref{dec-time1}) is given by $\operatorname{tr}\left(\rho_0^2 T_\nu T_\nu\right)=\frac{N^2-1}{2N} P_0$. The second summation is obtained using $T_\nu X T_\nu=\operatorname{Tr}(X) \id/2-X/(2N)$ \cite{Haber2021}, yielding $\operatorname{tr}\left(\rho_0 T_\nu \rho_0 T_\nu\right)=\frac{1}{2}\left(1-\frac{P_0}{N}\right)$. Combining these results leads to
\begin{equation}
    t_\mathrm{deco}=\frac{P_0}{\gamma\left(N P_0-1\right)}.
    \label{dec-time2}
\end{equation}

An alternative way to obtain the decoherence time is to expand the purity
\begin{equation}
P_t=\frac{1}{N}+\left(P_0-\frac{1}{N}\right)\mathrm{e}^{-\gamma N t},
\end{equation}
which follows directly from the solution of the Lindblad equation (see Eq. (7) in the Letter). Writing
\begin{equation}
P_t/P_0=1+\sum_{n=1}^{\infty}\frac{(-1)^n}{n!}\left(\frac{t}{t_{\mathrm{deco},n}}\right)^{n},
\qquad
t_{\mathrm{deco},n}
=\frac{1}{\gamma N}
\left(\frac{N P_0}{N P_0-1}\right)^{1/n}
\quad (n\ge1),
\end{equation}
makes explicit that all higher-order timescales are related. Thus, the coefficients are not independent, and throughout this work we identify the decoherence time with the leading term ($n=1$).

\subsection{c. The finite vanishing time $t_c$ of Wigner negativity}
We first derive the critical time $t_c$ at which the Wigner distribution becomes nonnegative, starting from a general initial state $\rho_0$. As shown in the main text, the Wigner function is given by (see Eq. (11) in the Letter, with $s=0$)
\begin{equation}
\label{Wigner_evolution}
    W^{(0)}_{\rho_0}(\Omega)= \frac{1-r_0}{N}+r_0 \,\langle\Omega|\rho_0|\Omega\rangle,
\qquad r_0=\sqrt{N+1},
\end{equation}
where $\Omega\in \mathbb{C}\mathrm{P}^{N-1}$ parametrizes the coherent-state manifold.

The initial Wigner function $W^{(0)}_{\rho_0}(\Omega)$ develops a negative region if and only if its
minimum satisfies
\begin{equation}
\label{negativity_condition}
    \min_{\Omega} W^{(0)}_{\rho_0}(\Omega)=\frac{1-r_0}{N}+r_0 \lambda_{\min} < 0,
\end{equation}
where 
\begin{equation}
\label{minexp_lmin}
    \lambda_{\min}=\min_{\Omega}\langle\Omega|\rho_0|\Omega\rangle,
\end{equation}
is the smallest eigenvalue of $\rho_0$. Therefore, Eq. (\ref{negativity_condition}) requires 
\begin{equation}
\label{pc}
    \lambda_{\min}<p_c=\frac{r_0-1}{Nr_0}=\frac{\sqrt{N+1}-1}{N\sqrt{N+1}},
\end{equation}
where $p_c$ plays the role of a critical Wigner negativity threshold. From the perspective of the initial state purity, this threshold is equivalent to the condition
\begin{equation}
    P_0>\frac{N}{N^2-1}.
    \label{P0negativity}
\end{equation}

The Wigner negativity disappears when $W^{(0)}_{\rho_t}(\Omega)$ reaches zero.
From Eqs. \eqref{Wigner_evolution} and \eqref{negativity_condition} we obtain the critical time 
\begin{equation}
\label{tc}
    t_c=\frac{2}{\gamma N}\,
\ln\!\Big[\sqrt{N+1}\,(1-N\lambda_{\min})\Big].
\end{equation}
For a pure initial state ($\lambda_{\min}=0$), Eq. \eqref{tc} reduces to 
\begin{equation}
    t_c=\frac{\ln(1+N)}{\gamma N}.
\label{tc2}
\end{equation}

\subsection{d. Ratio $t_c/t_{\mathrm{deco}}$ and the condition for which it is less than unity}
With Eqs. (\ref{dec-time2}) and (\ref{tc}), we obtain the ratio
\begin{equation}
   \frac{t_c}{t_\mathrm{deco}}= \frac{2(NP_0-1)}{NP_0}\ln\!\Big[\sqrt{N+1}\,(1-N\lambda_{\min})\Big].
   \label{ratio}
\end{equation}
For initial pure states ($\lambda_{\min}=0$), Eq. \eqref{ratio} reduces to 
\begin{equation}
    \frac{t_c}{t_\mathrm{deco}}=\ln (1+N)\left(1-\frac{1}{N}\right)\begin{cases}<1 & \text { for } N=2,3 \\ >1 & \text { for } N\geq 4\end{cases},
\end{equation}
which implies that for $N=2,3$, the critical time precedes the decoherence time. 
Particularly, in the large dimension limit ($N\to\infty$) and for a pure initial state, we find $t_c/t_{\mathrm{deco}}\propto \ln N$. This implies that, in this scenario, the conventional definition of the decoherence time remains appropriate, as the decoherence time is shorter than the critical time. The above analysis arises from the assumption of a pure initial state. A natural question is whether, for a given purity $P_0$, there exists an initial state $\rho_0$ such that $t_c/t_{\mathrm{deco}}<1$? The answer is affirmative. In the following, we provide a detailed proof.

For fixed purity $P_0$, the ratio (Eq. \eqref{ratio}) is minimized when $\lambda_{\min}$ is as large as possible. The maximum possible $\lambda_{\min}$ is attained by the spectrum 
\begin{equation}
\Big(1-(N-1)m,\underbrace{m,\dots,m}_{N-1~\mathrm{times}}\Big),
\end{equation}
which follows from standard Schur-convexity arguments. Thus,
\begin{equation}
\label{eq:lmin_max_SM}
  \lambda_{\min}^{\max}(P_0)=\frac{1-\sqrt{\frac{NP_0-1}{N-1}}}{N}.
\end{equation}
Substituting it into Eq. \eqref{ratio}, we obtain
\begin{equation}
\label{min_ratio}
  \left(\frac{t_c}{t_{\mathrm{deco}}}\right)_{\min}(P_0)=\frac{NP_0-1}{NP_0}\;
\ln\!\Bigg[\frac{(N+1)(NP_0-1)}{N-1}\Bigg].
\end{equation}
For each integer $N$, we define the boundary purity $P_0^b$ by setting Eq. \eqref{min_ratio}$=1$, namely,
\begin{equation}
    P_0^b=\frac{1}{N}\left(1+\frac{1}{W\left(\frac{N+1}{e(N-1)}\right)}\right), \quad N \geq 4,
\label{Pstar}
\end{equation}
where $W(\cdot)$ is the Lambert $W$ function (principal branch). Then, for $N /\left(N^2-1\right)<P_0<P_0^b$ there exist initial states with $t_c/t_{\mathrm{deco}}<1$, while for $P_0>P_0^b$ the ratio $t_c/t_{\mathrm{deco}}>1$ for all states with that purity.

Throughout this section, we focus on the Wigner function (the $s=0$ Stratonovich-Weyl representation), as it is the most widely used quasiprobability function. 
For other Stratonovich-Weyl quasiprobability functions, one may similarly define an $s$-dependent finite time at which the corresponding phase-space distribution becomes nonnegative. 
It is worth noting that our single-shot positivity result in the main text is universal across the Stratonovich-Weyl family, in the sense that one round of the coherent-state POVM removes quasiprobability negativity for any choice of Stratonovich-Weyl kernel.

\section{IV. Derivation of Eq. (13) in the main text}
The spectral decomposition of initial state is expressed as $\rho_0=\sum_{j=1}^N \lambda_j |e_j\rangle\!\langle e_j|$
with the eigenvalues $\boldsymbol \lambda=(\lambda_1,\ldots,\lambda_N)$ and $\lambda_1\le\cdots\le\lambda_N$. Let $\boldsymbol{x}=(x_1,\ldots,x_N)$ with $x_j=|\langle e_j|\Omega\rangle|^2$ ($x\ge0$ and $\sum_j x_j=1$). Under coherent-state sampling, $\boldsymbol x \sim \mathrm{Unif}(\Delta_{N-1})$, which denotes the uniform (Lebesgue) distribution on the set 
$\Delta_{N-1}=\{\boldsymbol{x}\in\mathbb{R}^N:\, x_j\ge 0,\ \sum_{j=1}^N x_j=1\}$, i.e., the probability measure that assigns equal weight to all points in this $(N\!-\!1)$-dimensional simplex.

Define the random variable
\begin{equation}
  p=\langle\Omega|\rho_0|\Omega\rangle=\boldsymbol \lambda\!\cdot\!\boldsymbol{x}=\sum_{j=1}^N \lambda_j x_j.
\label{S1}
\end{equation}
Denote by $f_p$ the probability density function (pdf) of $p$ and by $F_p$ its cumulative distribution function (cdf), $F_p(u)=\Pr[p\le u]$ (i.e., $\mathcal P(\rho_0)$ in the main text). For the negativity threshold we use $u=p_c=(r_s-1)/(N r_s)$ with $r_s=(N+1)^{(1+s)/2}$. Then, the Laplace transform of $f_p(u)=F_p'(u)$ reads
\begin{equation}
  \mathcal L\{f_p\}(\zeta)=\int_0^\infty e^{-\zeta u} f_p(u) \mathrm d u
=\int_{\Delta_{N-1}} e^{-\zeta \boldsymbol \lambda\!\cdot\!\boldsymbol{x}}\mathrm d \boldsymbol{x}.
\label{A1}
\end{equation}
In terms of the truncated-power form of a univariate B-spline (see, e.g., Problem-IX10 in Ref. \cite{deBoor2001}), Eq. (\ref{A1}) yields, for pairwise distinct $\{\lambda_j\}$
\begin{equation}
  \int_{\Delta_{N-1}} e^{-\zeta \boldsymbol \lambda\!\cdot\!\boldsymbol{x}}\mathrm d \boldsymbol{x}
=\frac{1}{(N-1)!}\,\frac{1}{\zeta^{\,N-1}}
\sum_{j=1}^{N}\frac{e^{-\zeta\lambda_j}}{\displaystyle\prod_{k\ne j}(\lambda_j-\lambda_k)}.
\label{S3}
\end{equation}
The inverse Laplace transform of Eq. (\ref{S3}) is given by 
\begin{equation}
  f_p(u)=\frac{1}{(N-2)!} \sum_{j=1}^N \frac{\left(u-\lambda_j\right)_{+}^{N-2}}{\prod_{k \neq j}\left(\lambda_j-\lambda_k\right)},
  \label{S4}
\end{equation}
where $(\cdot)_{+}$ means the positive part. Integrating $f_p(u)$ from $0$ to $p_c$, yields
\begin{equation}
  F_p(u)=\int_0^{p_c} f_p(u)\mathrm d u
=\frac{1}{(N-1)!}\sum_{j=1}^{N}
\frac{(p_c-\lambda_j)_+^{\,N-1}}{\displaystyle\prod_{k\ne j}(\lambda_j-\lambda_k)}.
\label{S5}
\end{equation}
Since $(p_c-\lambda_j)_+^{N-1}=0$ whenever $\lambda_j\ge p_c$, this is equivalently
\begin{equation}
  \mathcal P(\rho_0)=F_p(u)
=\frac{1}{(N-1)!}\sum_{\lambda_j<p_c}
\frac{(p_c-\lambda_j)^{N-1}}{\displaystyle\prod_{k \ne j}(\lambda_j-\lambda_k)},
\label{S6}
\end{equation}
which is Eq. (13) in the main text. For degenerate eigenvalues, Eq. \eqref{S6} is obtained by taking the appropriate continuous limit.

\section{V. Experimental Proposal}
The theory developed in the Letter can be implemented on state-of-the-art experimental platforms, such as superconducting circuits \cite{Clarke2008,Huang2020,Kjaergaard2020,Jiang2025}. Here, we briefly outline a feasible proposal for experimental verification, consisting of the following three steps.

(i) Haar measure on $\mathbb{C}\mathrm{P}^{N-1}$. Since $\mathrm{d}\mu(\Omega) = \prod_{k=1}^{N-1} [\,p_k(\theta_k)\,\mathrm{d}\theta_k \times \mathrm{d}\phi_k / (2\pi)\,]$, we can sample $\Omega$ for $E(\Omega)$ by drawing each $\theta_k$ from $p_k(\theta_k) = (N - k)\,\sin^{2(N - k) - 1}\!\left(\tfrac{\theta_k}{2}\right)\!\cos\!\left(\tfrac{\theta_k}{2}\right)$, and each $\phi_k$ uniformly from $[0, 2\pi)$.

\begin{figure}[t]
	\centering
	\includegraphics[width=9.0cm]{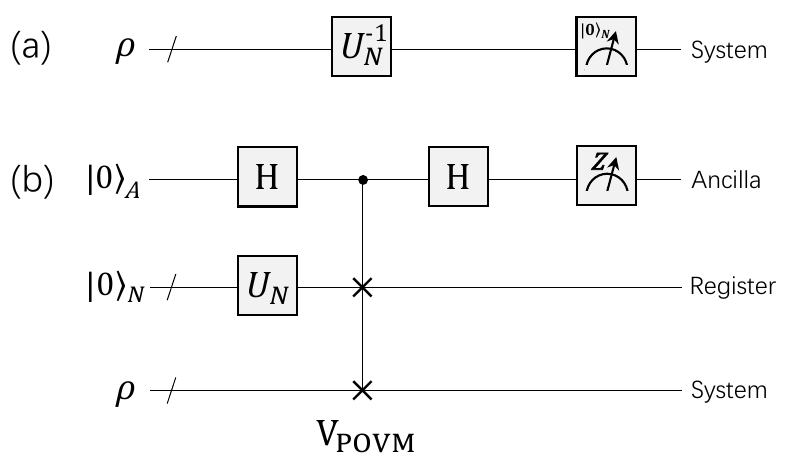}
	\caption{\textbf{Schematic quantum circuits for implementing $N$-level coherent-state POVMs.} (a) Direct measurement protocol via application of the inverse unitary $U_N^{-1}$, i.e., $\operatorname{Tr}[U_N^{-1} \rho U_N |0\rangle_N \langle 0|] = \operatorname{Tr}[\rho E(\Omega)] / N$. (b) Ancilla-assisted protocol for detecting the POVM using both an ancilla and a reference register. The key step involves a controlled-SWAP gate $V_{\text{POVM}}$ between the system and the register. The outcome is obtained by measuring the ancilla in the $Z$ basis, yielding $\langle Z_A \rangle = \operatorname{Tr}[\rho E(\Omega)] / N$, which matches the result from (a) with unit success probability at the cost of increased resource requirements. }
	\label{fig_schematic}
\end{figure}

(ii) $N$-level coherent-state POVMs. The central task is to evaluate $\operatorname{Tr}[\rho E(\Omega)]$, which can be implemented either probabilistically using an ancilla-free strategy or deterministically with an ancilla-assisted strategy. For the former, since $|\Omega\rangle = U_N(\Omega) |0\rangle_N$, this can be realized by applying the inverse unitary $U_N^{-1}(\Omega)$ to $\rho$, performing a measurement in the computational basis, and post-selecting the outcome $|0\rangle_N$ (see Fig. \ref{fig_schematic}(a)). The advantage of this approach is that it requires no additional auxiliary qubits, although the success probability is limited. 

For the latter approach, an ancilla qubit and a reference register are required, as illustrated in Fig. \ref{fig_schematic}(b). A single-qubit ancilla is initialized in $|0\rangle_A$ and subjected to a Hadamard gate $\mathrm{H}$. We then sample $\Omega$ and prepare a second $N$-level register in the state $|\Omega\rangle$, which leads to $\varrho_1=\frac{1}{2} \sum_{x, y \in\{0,1\}}|x\rangle\left\langle\left. y\right|_A \otimes  \mid \Omega\right\rangle\left\langle\left.\Omega\right|_R\right. \otimes\rho$. The following key step is the application of a controlled-SWAP gate, $V_{\text{POVM}} = |0\rangle\langle 0|_A \otimes \mathbf{1}_{RS} + |1\rangle\langle 1|_A \otimes U_{\text{SWAP}} $, which conditionally exchanges the contents of the reference register $R$ and the system $S$, and leads to $\varrho_2=V_{\mathrm{POVM}} \varrho_1 V_{\mathrm{POVM}}^{\dagger}$. A second Hadamard gate is then applied to the ancilla, followed by a measurement in the Pauli-$Z$ basis. This yields the expectation value $\langle Z_A \rangle =\operatorname{Tr}[\left(Z_A \otimes \mathbf{1}_{RS}\right)\left(\mathrm{H} \otimes \mathbf{1}_{RS}\right) \varrho_2\left(\mathrm{H} \otimes \mathbf{1}_{RS}\right)^{\dagger}]= \operatorname{Tr}[\rho E(\Omega)]/N$, realizing the desired POVM. The advantage of this method is that it is deterministic and well-suited for low-dimensional systems, where the required ancilla and reference register resources remain feasible.

For both implementations, the circuit depth scales linearly with the Hilbert-space dimension, $\Theta(N)$ \cite{LCU2012}. 
Let the fidelity of a single two-qubit gate be $f$ (with infidelity $r = 1 - f$) and the gate time be $\tau$. 
The circuit consists of $L$ two-qubit layers, with $L \simeq \kappa N$, where $\kappa$ is a circuit-dependent constant. 
The total gate time of the circuit is therefore $T_{\mathrm{tot}} \simeq L \tau$, and the total fidelity is $F_{\mathrm{tot}} \simeq (1 - r)^L \simeq e^{-rL}$. 
If the minimum acceptable total fidelity is $F_{\min}$, the maximum implementable system size satisfies
\begin{equation}
  N \lesssim \min\!\left\{\frac{-\ln F_{\mathrm{min}}}{\kappa r}, \frac{T_{2}}{\kappa \tau}\right\}.
  \label{Ngu}
\end{equation}
Current superconducting platforms report two-qubit gate fidelities of $99.7$–$99.9\%$, gate times of $20$–$160$ ns, and coherence times in the range $T_{2} \sim 0.3$–$0.5$ ms for transmons and $\gtrsim 1$ ms for fluxonium \cite{Clarke2008,Huang2020,Kjaergaard2020,Jiang2025}. 

For a rather conservative estimate, we assume $F_{\min}=95\%$, $T_2=0.3~\mathrm{ms}$, $\kappa=1.5$, $r=0.2\%$, and $\tau=100 \mathrm{ns}$. Under these assumptions, our estimates suggest that the protocol remains experimentally feasible up to approximately $N\lesssim17$. Using the same bound but adopting improved superconducting parameters already quoted in the manuscript, including two-qubit fidelities up to $f\simeq99.9\%$ and $\kappa=1$, Eq. \eqref{Ngu} suggests feasibility up to $N\lesssim51$. In both cases, these system sizes are sufficient to observe the main phenomena predicted in this Letter.

(iii) Measurement in phase space. In experiments, one may obtain the density matrix after performing the above POVM and directly compute the phase-space function. Alternatively, the function can be measured digitally by constructing the SW kernel via the linear-combination-of-unitaries technique \cite{LCU2012}.

\section{VI. Heralded resource extraction from unital depolarizing noise}\label{IV}
We may raise a natural operational question: although the depolarizing channel is unital and typically viewed as purely detrimental, can one nevertheless obtain coherent states (or other useful resources) from such unital noise?
Since Eq. \eqref{unital} is a depolarizing (hence unital and mixing) channel, it is purity-contracting \cite{LIDAR200682}. This implies that no deterministic protocol based solely on the depolarizing map can output a pure coherent state from a generic input. Therefore, if the depolarizing noise is treated as a black box without access to its underlying measurement record, coherent-state preparation from the noise alone is impossible.

The above impossibility result relies crucially on discarding the measurement record. In contrast, if one has access to an unraveling of the depolarizing channel in terms of the coherent-state POVM, then each individual run produces a pure state conditioned on the observed outcome $\Omega$, i.e., a heralded coherent-state preparation. In what follows, we outline the protocol.

Introduce an environment register with orthonormal pointer states $\{\ket{\Omega}_\mathrm{E}\}$.
We emphasize that $\{\ket{\Omega}_\mathrm{E}\}$ is merely an orthonormal pointer basis of the environment register serving as a classical record. The label $\Omega$ here is only an outcome index and does not imply that the record states are coherent states.
Because the coherent-state POVM has a continuous outcome space, the canonical dilation employs an infinite-dimensional record register
with generalized pointer states $\ket{\Omega}_\mathrm{E}$ satisfying ${}_\mathrm{E}\langle \Omega|\Omega'\rangle_\mathrm{E}=\delta(\Omega,\Omega') $, where $\delta(\Omega,\Omega')$ denotes the Dirac delta distribution with respect to the measure $\text{d}\mu(\Omega)$ on the outcome manifold $\mathcal M$, i.e.,
\begin{equation}
  \int_{\mathcal M} \text{d}\mu(\Omega)\,\delta(\Omega,\Omega')\,f(\Omega)=f(\Omega') .
\end{equation}
In practice, one may discretize $\Omega$ to obtain an approximate implementation with a finite-dimensional register. Now, let $U_{\text{SE}}$ denote a joint unitary acting on the system and an environment register initialized in $\ket{0}_\text{E}$. For any $\ket{\psi}_\text{S}\in\mathcal H_\text{S}$,
\begin{equation}
  U_\text{SE}\big(\ket{\psi}_\text{S}\otimes\ket{0}_\text{E}\big)
  =\int_{\mathcal M} \text{d}\mu(\Omega)\,\big(M(\Omega)\ket{\psi}_\text{S}\big)\otimes\ket{\Omega}_\text{E}.
\end{equation}
Consider an arbitrary initial system state $\rho$, with the environment initially prepared in $\ket{0}_{\mathrm E}$ (i.e., $\rho_\text{SE}=\rho \otimes \ket{0}_\text{E}\bra{0}$). Given the joint state after the interaction, it is convenient to express it directly in terms of $U_{\text{SE}}$
\begin{equation}
  \rho'_{SE}=U_\text{SE}\big(\rho\otimes\ket{0}\!\bra{0}_\text{E}\big)U_\text{SE}^\dagger.
\end{equation}
We note that discarding the environment record corresponds to tracing out the environment, which recovers the unconditional channel 
\begin{equation}
  \mathcal E(\rho)
  =\text{Tr}_\text{E}[\rho_{SE}']
  =\int_{\mathcal M} \text{d}\mu(\Omega)\,M(\Omega)\rho M(\Omega)^\dagger .
  \label{epsilonrho}
\end{equation}
This unconditional case is not our focus here. Instead, we consider a measurement of the environment register that generates the classical record $\Omega$, modeled as a projective measurement in the pointer basis $\{\ket{\Omega}_\mathrm{E}\}$. The probability density for observing the outcome $\Omega$ is then
\begin{equation}         p(\Omega)=\operatorname{Tr}_{\mathrm{SE}}\left[\rho_{\mathrm{SE}}^{\prime}\left(\mathbf{1}_N \otimes|\Omega\rangle_{\mathrm{E}}\langle\Omega|\right)\right]=\operatorname{Tr}_{\mathrm{S}}\left[\rho M(\Omega)^{\dagger} M(\Omega)\right] .
    \label{pOmega} 
\end{equation}
Moreover, the unnormalized conditional (post-measurement) system state is
\begin{equation}
    \tilde{\rho}_{\Omega}=\text{Tr}_\text E\left[\rho_\text{SE}^{\prime}\left(\mathbf{1}_{N} \otimes|\Omega\rangle_{\mathrm{E}}\left\langle\left.\Omega\right|\right)\right]=M(\Omega) \rho M(\Omega)^{\dagger} .\right.
    \label{rhoOmega}
\end{equation}
Normalizing by $p(\Omega)$ yields the conditional system state
\begin{equation}
  \rho_\Omega
  =\frac{\tilde{\rho}_\Omega}{p(\Omega)}
  =\frac{M(\Omega)\rho M(\Omega)^\dagger}{\text{Tr}_\text{S}\!\left[\rho\,M(\Omega)^\dagger M(\Omega)\right]} .
\end{equation}
For $M(\Omega)=\sqrt{N}\ket{\Omega}\!\bra{\Omega}$, one has
$M(\Omega)\rho M(\Omega)^\dagger = N\,\langle\Omega|\rho|\Omega\rangle\,\ket{\Omega}\!\bra{\Omega}$ and
$p(\Omega)=N\,\langle\Omega|\rho|\Omega\rangle$, hence $\rho_\Omega=\ket{\Omega}\!\bra{\Omega}$ for every outcome $\Omega$.

\begin{figure}[t]
	\centering
	\includegraphics[width=8.5cm]{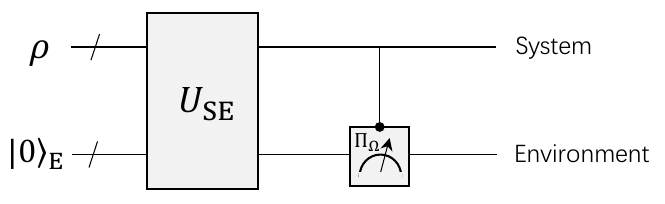}
	\caption{\textbf{Schematic quantum circuit for heralded resource extraction from unital depolarizing noise.} 
A joint unitary $U_\text{SE}$ correlates the system $\rho$ with an environment register (initialized in $\ket{0}_\text{E}$), followed by an environment measurement $\{\Pi_\Omega\}=\ket{\Omega}_\mathrm{E} \bra{\Omega}$ that generates a classical record $\Omega$. The vertical link indicates correlation rather than a controlled operation. Retaining $\Omega$ realizes a heralded coherent-state preparation on the system (Eq. \eqref{rhoOmega}), whereas discarding $\Omega$ recovers the unconditional depolarizing channel $\mathcal{E}$ (Eq. \eqref{epsilonrho}).
}
	\label{fig_resource}
\end{figure}

While the depolarizing channel is unital and therefore cannot generate purity or coherence resources at the unconditional (ensemble-averaged) level, monitoring the same dynamics through an unraveling compatible with the coherent-state POVM yields pure quantum trajectories, each corresponding to a coherent state. In this sense, resources can be extracted from unital noise when the classical measurement record is available. This trajectory level perspective provides a concrete operational way for converting isotropic depolarization into a heralded resource generator in finite-dimensional platforms.

Finally, we note that operationally, the nontrivial requirement is the ability to implement (or effectively engineer) a system-environment interaction and an environment measurement scheme whose induced instrument coincides with $\{M(\Omega)\}$.
Although such a dilation exists mathematically, its physical realization is model and platform dependent.

\end{document}